\documentclass[conference]{IEEEtran}
\IEEEoverridecommandlockouts
\usepackage{cite}
\usepackage{amsmath,amssymb,amsfonts}
\usepackage{algorithmic}
\usepackage{graphicx}
\usepackage{textcomp}
\usepackage{xcolor}
\usepackage{subfigure}
\usepackage[strings]{underscore} % it fix the anoying error in .output.bbl
\def\BibTeX{{\rm B\kern-.05em{\sc i\kern-.025em b}\kern-.08em
    T\kern-.1667em\lower.7ex\hbox{E}\kern-.125emX}}

\begin{document}

\title{Learning the Frequency Dynamics of the Power System Using Higher-order Dynamic Mode Decomposition\\
\thanks{This research is funded by Sino-German (CSC-DAAD) Postdoc Scholarship Program(91870333) and the Helmholtz Association's Initiative and Networking Fund through Helmholtz AI.}
}
\author{
\IEEEauthorblockN{Xiao Li}
\IEEEauthorblockA{\textit{\parbox{5cm}{\centering Institute for Automation and Applied Informatics\\
Karlsruhe Institute of Technology}} \\
Karlsruhe, Germany \\
xiao.li@kit.edu}
\and
\IEEEauthorblockN{Xinyi Wen}
\IEEEauthorblockA{\textit{\parbox{5cm}{\centering Institute for Automation and Applied Informatics\\
Karlsruhe Institute of Technology}} \\
Karlsruhe, Germany \\
xinyi.wen@kit.edu}
\and
\IEEEauthorblockN{Benjamin Schäfer}
\IEEEauthorblockA{\textit{\parbox{5cm}{\centering Institute for Automation and Applied Informatics\\
Karlsruhe Institute of Technology}} \\
Karlsruhe, Germany \\
benjamin.schaefer@kit.edu}
}

\maketitle

\begin{abstract}
The increasing penetration of renewable energy sources, characterised by low inertia and intermittent disturbances, presents substantial challenges to power system stability. As critical indicators of system stability, frequency dynamics and associated oscillatory phenomena have attracted significant research attention. While existing studies predominantly employ linearized models, our findings demonstrate that linear approximations exhibit considerable errors when predicting frequency oscillation dynamics across multiple time scales, thus necessitating the incorporation of nonlinear characteristics. This paper proposes a data-driven approach based on higher-order dynamical mode decomposition (HODMD) for learning frequency dynamics. The proposed method offers distinct advantages over alternative nonlinear methods, including no prior knowledge required, adaptability to high-dimensional systems, and robust performance. Furthermore, HODMD demonstrates superior capability in capturing system-wide spatio-temporal modes, successfully identifying modal behaviour that remains undetectable through standard Dynamic Mode Decomposition techniques. The efficacy of the proposed methodology is validated through comprehensive case studies on both IEEE 14-bus and WECC systems.
\end{abstract}

\begin{IEEEkeywords}
Frequency oscillation, HODMD, patio-temporal mode, data-driven method, model-free method
\end{IEEEkeywords}

\section{Introduction}

The power system's reliable and stable operation is fundamental to society, necessitating effective monitoring and modelling of system operation, like frequency measurement, for control implementation. Traditional equation-based analytical methods have shown limitations in accurately characterizing power system frequency dynamics \cite{aththanayake_power_2024}. With increasing renewable energy integration and power fluctuations \cite{schafer_non-gaussian_2018}, data-driven, model-free approaches demonstrate greater adaptability to future grid requirements \cite{anvari_data-driven_2022}. 

Recent advances in data-driven methods, coupled with mature wide-area measurement technologies, have shown promise in directly learning power system dynamics from data\cite{gong_data-driven_2023, gorjao_data-driven_2020}. Among them, dynamical mode decomposition (DMD) \cite{barocio_dynamic_2015, alassaf_randomized_2021}, sparse identification of nonlinear dynamical systems (SINDy)\cite{wen_identifying_2024}, Gaussian processes Regression (GPR) \cite{heinonen_learning_2018}, and neural network-based approaches \cite{zhi_learning_2022} have received considerable attention. DMD provides a robust approach to dynamics analysis by decomposing complex systems into dynamic modes that reveal linear or near-linear behaviour within spatio-temporal structures. With increasing data size or dimensional subspace of the space of observables, DMD converges to the Koopman operator, providing a theoretical basis for system identification \cite{korda_convergence_2018}. Conversely, SINDy uses sparse regression to discover the governing equations directly from the data, capturing nonlinear behaviour while identifying the minimum terms necessary to describe the system, which is particularly suitable for systems with sparse but unknown structures\cite{brunton_discovering_2016}.

Nevertheless, several challenges arise when applying these methods to learn the frequency dynamics of power systems. In this paper, we are concerned with power system frequency variations under disturbances, especially frequency oscillations. 
The first challenge is that the frequency dynamics of the power system exhibit nonlinear characteristics even in small deviations from the equilibrium point, and thus linear methods do not work well, such as the standard DMD method.
The second challenge is that strongly nonlinear methods often require a very strong selection of the appropriate feature function, but the selection of the feature function often limits the applicability of the method, which can make the effect of the method remain localized, e.g. Extended DMD (EDMD) \cite{williams_datadriven_2015}, Linear and Nonlinear Disambiguation Optimization (LANDO) \cite{baddoo_kernel_2022}, and SINDy. We also compare these methods in this paper and find that they are no better than higher-order DMD (HODMD) \cite{vega_higher_2020}.
The third challenge is that when studying the frequency-dynamic spatio-temporal characteristics of interconnected power systems, the dimensions of the system are so large that techniques such as SINDy, GPR, etc. are difficult to apply.  

Our study employs the HODMD to learn the frequency dynamics of the power system and analyse the global and local oscillation modes. The remainder of the paper is structured as follows: The first part describes the basic characteristics of power system frequency dynamics, while the second part presents the basic idea of the HODMD algorithm and the steps to learn frequency dynamics from data using the proposed method. Finally, we validate our method on two test systems.

\section{The frequency dynamics of power system}

The power system is a complex network, each node is a dynamic device that depends on the node voltage (i.e., the bus voltage), and the nodes are connected by transmission lines or directly (e.g., devices connected to the same bus). In this paper, the frequency of the power system refer to the  angular frequency of each generator rotor. It has an explicit algebraic relationship with the frequency of the bus voltage\cite{zhao_robust_2018}, so we can extend our methods to renewable energy sources easily. 
These dynamics of the rotor of the generator can be represented by the swing equations, as Eq.(\ref{eqs:single_machine_frequency}):
\begin{equation}
 \left\{
 \begin{aligned} 
    \frac{\mathrm{d}\delta_i}{\mathrm{d}t}  &=  \omega_0(\omega_i - 1) \\
    J_i\frac{\mathrm{d}\omega_i}{\mathrm{d}t} &= P_{\mathrm{m},i}- P_{\mathrm{e},i} - D_i(\omega_i-\omega_0), 
    i= 1,2,... N 
\end{aligned}
\label{eqs:single_machine_frequency}
\right.   
\end{equation}
where the subscript $i$ denotes the $i$-th generator, with $J_i$ and $D_i$ representing the equivalent inertia and damping, respectively. $P_{\mathrm{m},i}$ and $P_{\mathrm{e},i}$ denotes the mechanical turbine power and the electrical generator power, respectively. $\delta_i$ denotes rotor angle. The normalized angular frequency $\omega_i$ is related to the nominal angular velocity $\omega_0=2\pi f_n$, where $f_n$ is the nominal frequency. 

On the network side of the power system, the power balance equation of $P_{\mathrm{e},i}$ is the relationship connecting the nodes of the complex system. Thus, the frequency dynamics of the different generator rotors are coupled to each other through the network. Due to the limited range of frequency deviation, the model for the network side is often approximated by the linear model, as shown in Eq.(\ref{eqs:network}) \cite{gao_common-mode_2022}.
\begin{equation}
    \mathrm{d} \boldsymbol{P}_{\mathrm{e}} = \boldsymbol{L}\mathrm{d} \boldsymbol{\delta}\label{eqs:network}
\end{equation}
where $\boldsymbol{P}_{\mathrm{e}}$ and $\boldsymbol{\delta}$ denote the vector of the ${P}_{\mathrm{e}, i}$ and $\delta_{i}$, respectively, and $\boldsymbol{L}$ denotes a constant matrix including the network topology, parameters and steady state.

As the linear network Equation (\ref{eqs:network}), the nonlinear characteristics of the frequency dynamics of the power system are mainly reflected in the swing Equation (\ref{eqs:single_machine_frequency}). Therefore, we can approximate the frequency dynamics of the power system by using the rotor angle and angular frequency of the generator.

\section{Methodologies}

\subsection{Preliminary of HODMD}
In this section, we briefly introduce the method Higher-order Dynamic Mode Decomposition (HODMD)\cite{vega_higher_2020}, which is an extension of the standard DMD method. The standard DMD method assumes that the dynamics of the system can be approximated by the linear Koopman eigenfunction so that its state variables $\boldsymbol{x}(t_k)$ can be represented in the form of Eq.(\ref{eqs:dmd_assumption}).
\begin{equation}
    \boldsymbol{x}(t_k) \simeq \sum_{m=1}^{r} \boldsymbol{u}_{m} a_m \lambda_m ^{(k-1)\Delta t} \quad \text {for } k=1,...,T
\label{eqs:dmd_assumption}
\end{equation}
where $\boldsymbol{u}_{m}$ is normalized $m$-th spatial modes, and $a_m > 0$ is the mode amplitudes, and $\lambda_m$ denotes the eigenvalues of modes. $r$ is the total number of modes and also the truncated rank of singular value decomposition (SVD) in DMD chosen by certain criteria.

But even for systems of the form Eq.(\ref{eqs:dmd_assumption}), the standard DMD method does not always give correct results. Since the space
of observables is equal to or less than that of state variables in standard DMD, $r$ won't be more than the number of state variables, which is not a problem in a linear system. However, due to the weakly nonlinear nature of the system, the good approximation of the Koopman modes of the system dynamics will be greater than the number of state variables when the system state variables are expanded according to the form of Eq.(\ref{eqs:dmd_assumption}). The HODMD method solves this problem by proposing exactly the delayed embedding. Eq.(\ref{eqs:hodmd}) is the essence of the HODMD method. This acts like a higher-order difference equation, allowing it to handle weakly nonlinear dynamics.
\begin{equation}
    \boldsymbol{x}_{k+1}=\boldsymbol{R}_{1} \boldsymbol{x}_{k}+\boldsymbol{R}_{2} \boldsymbol{x}_{k-1}+\cdots+\boldsymbol{R}_{d} \boldsymbol{x}_{k-d+1}
    \label{eqs:hodmd}
\end{equation}
where $k=1,\cdots,T-1$, $d\geq1$ and $\boldsymbol{R}_{i}$ is the constant number coefficient matrix. When $d=1$, Eq.(\ref{eqs:hodmd}) downgrades to standard DMD.

We rewrite the delay term as a standard DMD, as shown in Eq.(\ref{eqs:modified_snapshots}):
\begin{equation}
    \tilde{\boldsymbol{x}}_{k+1}=\tilde{\boldsymbol{R}} \tilde{\boldsymbol{x}}_{k}
\label{eqs:modified_snapshots}
\end{equation}
where
\begin{equation*}
\tilde{\boldsymbol{x}}_{k} \equiv\left[\begin{array}{c}
    \boldsymbol{x}_{k} \\
    \boldsymbol{x}_{k+1} \\
    \ldots \\
    \boldsymbol{x}_{k+d-2} \\
    \boldsymbol{x}_{k+d-1}
    \end{array}\right], \quad \tilde{\boldsymbol{R}} \equiv\left[\begin{array}{ccccc}
    \mathbf{0} & \mathbf{I} & \ldots & \mathbf{0}  &  \mathbf{0} \\
    \mathbf{0} & \mathbf{0} & \ldots & \mathbf{0}  &  \mathbf{0} \\
    \ldots     & \ldots     & \ldots & \ldots      &  \ldots \\
    \mathbf{0} & \mathbf{0} & \ldots & \mathbf{0}  &\mathbf{I} \\
    \boldsymbol{R}_{1} & \boldsymbol{R}_{2}  & \ldots & \boldsymbol{R}_{d-1} & \boldsymbol{R}_{d}
\end{array}\right]
\end{equation*}

\subsection{Learning the frequency dynamics using HODMD}

Based on the time series data of one scenario of the generators' rotor angles $\boldsymbol{\delta}(t)$ and their normalized angular frequency $\boldsymbol{\omega}(t)$ on time $\boldsymbol{t} = [t_1,t_2,...t_T]$, we have the snapshot of $k$-th time step $\boldsymbol{x}_k = [\boldsymbol{x}^{(1)}(t_k),\boldsymbol{x}^{(2)}(t_k),...,\boldsymbol{x}^{(N)}(t_k)]^T$, where $\boldsymbol{x}^{(i)}(t_k) = [\omega_i(t_k),\delta_i(t_k)]$, $i =1,2,\cdots,N$, $k = 1,2,\cdots,T$.  
After that, the $d$-order delay term in the form of Eq.(\ref{eqs:modified_snapshots}) is obtained as $\tilde{\boldsymbol{X}}_{1}$.
\begin{equation}
   \tilde{\boldsymbol{X}}_{1} \equiv \left[\begin{array}{cccc}
    \omega_1(t_1) & \omega_1(t_2) & \ldots &  \omega_1(t_{T-d+1}) \\
    \delta_1(t_1) & \delta_1(t_2) & \ldots &  \delta_1(t_{T-d+1}) \\
    \ldots     & \ldots     & \ldots & \ldots   \\
    \omega_{N}(t_{1}) & \omega_{N}(t_{2}) & \ldots & \omega_{N}(t_{T-d+1}) \\
    \delta_{N}(t_{1}) & \delta_{N}(t_{2})  & \ldots & \delta_{N}(t_{T-d+1})\\
    \omega_1(t_2) & \omega_1(t_3) & \ldots &  \omega_1(t_{T-d+2}) \\
    \delta_1(t_2) & \delta_1(t_3) & \ldots &  \delta_1(t_{T-d+2}) \\
    \ldots     & \ldots     & \ldots & \ldots   \\
    \omega_{N}(t_{d}) & \omega_{N}(t_{d+1}) & \ldots & \omega_{N}(t_{T}) \\
    \delta_{N}(t_{d}) & \delta_{N}(t_{d+1})  & \ldots & \delta_{N}(t_{T})
\end{array}\right]
\end{equation}
Thus, each column of the modified snapshot matrix $\tilde{\boldsymbol{X}}_{1}$ is a snapshot of one moment in time, which is as many as $(2\cdot N\cdot d)$, with a total of $T-d+1$ time steps.  Based on the steps of the standard DMD method, the modified snapshot matrix $\tilde{\boldsymbol{X}}_{1}$ is first split into a matrix $\tilde{\boldsymbol{X}}_{\mathrm{X},1}$ from the first  $T-d$ time steps and a matrix $\tilde{\boldsymbol{X}}_{\mathrm{Y},1}$ from the last $T-d$ time steps. 

Due to the ability of the DMD method to decompose spatio-temporal modes, superimposing different perturbation scenarios in the snapshots can increase the accuracy of the results. 
But for $Q$ different perturbation scenarios, we can get $Q$ such snapshot matrices.
%and if we process them separately then we get $Q$ results. In this way, we will face the problem that in the end we choose which result is better. 
Instead, we synthesize the snapshots of multiple trajectories into one matrix, dealing with the temporal discontinuities between different trajectories as follows: First $\tilde{\boldsymbol{X}}_{\mathrm{X},i}$ and $\tilde{\boldsymbol{X}}_{\mathrm{Y},i}$, where $i=1,2,...,Q$, are computed separately for each perturbed scene, and then $\tilde{\boldsymbol{X}}_{\mathrm{X},i}$ is synthesised into a matrix in the time dimension, as well as $\tilde{\boldsymbol{X}}_{\mathrm{Y},i}$ is also synthesised into a matrix in the time dimension. This allows for a low-dimensional approximation to be computed.

\begin{equation}
    \tilde{\boldsymbol{X}}_\mathrm{Y} = \tilde{\boldsymbol{R}}\tilde{\boldsymbol{X}}_\mathrm{X}
\end{equation}
where 
$\tilde{\boldsymbol{X}}_\mathrm{X} = [\tilde{\boldsymbol{X}}_{\mathrm{X},1},\cdots,\tilde{\boldsymbol{X}}_{\mathrm{X},Q}]$, 
$\tilde{\boldsymbol{X}}_\mathrm{Y} = [\tilde{\boldsymbol{X}}_{\mathrm{X},1},\cdots,\tilde{\boldsymbol{X}}_{\mathrm{X},Q}]$.

After these processes above, the standard DMD steps follow\cite{kutz_dynamic_2016}:  
\begin{enumerate}
    \item Apply SVD $\tilde{\boldsymbol{X}}_\mathrm{X} \approx \boldsymbol{U}\boldsymbol{\Sigma} \boldsymbol{V}^H$, where $\boldsymbol{U} \in C^{2Nd\times r}$ is a unitary matrix, $\boldsymbol{\Sigma} \in C ^{r\times r}$ is a diagonal matrix with singular values, $\boldsymbol{V}^H$ is the conjugate transpose of unitary matrix $\boldsymbol{V} \in C^{(T_\text{total}-Qd)\times r}$, $r$ is the truncated rank, and $T_\text{total}$ is the total time steps of all the trajectories.
    \item Compute $\boldsymbol{K} = \boldsymbol{U}^H \tilde{\boldsymbol{X}}_\mathrm{Y}\boldsymbol{V}\Sigma^{-1}$; use it as a low-rank $(r \times r)$ approximation of $\tilde{\boldsymbol{R}}$.
    \item Compute eigendecomposition of $\boldsymbol{K}: \boldsymbol{K}\boldsymbol{W} = \boldsymbol{W}\boldsymbol{\Lambda}$, where $\boldsymbol{\Lambda} = (\lambda_m)$ are eigenvalues and columns of $\boldsymbol{W}$ are the corresponding eigenvectors. 
    \item Eigenvalues of $\tilde{\boldsymbol{R}}$ can be approximated by $\boldsymbol{\Lambda}$ with corresponding eigenvectors in the columns of $\boldsymbol{\Phi} = \boldsymbol{U}\boldsymbol{W}$.
\end{enumerate}

\subsection{The spatio-temporal modes analysis}

Using the Koopman eigenvalues and modes of the low-rank approximation, we reconstruct the frequency dynamics from Eq.(\ref{eqs:hodmd_reconstructure}) and then analyse its spatio-temporal modes.

\begin{equation}
    \boldsymbol{x}_{k+d-1}=\boldsymbol{L}\boldsymbol{\Phi} \boldsymbol{\Lambda}^{k-1} \boldsymbol{\Phi}^{-1}  \tilde{\boldsymbol{x}}_{1} , \quad k=1,2,\cdots
    \label{eqs:hodmd_reconstructure}
\end{equation}
where $\boldsymbol{L}$ is the reconstruction matrix and the first $2N$ columns are the identity matrix, $\boldsymbol{L} = [\mathbf{I},\mathbf{0},\cdots,\mathbf{0}]\in R^{2N \times 2Nd}$; $\boldsymbol{x}_{k+d-1}$ and $\tilde{\boldsymbol{x}}_{1}$ have the same meaning in Eq.(\ref{eqs:modified_snapshots}).

From the Eq.(\ref{eqs:dmd_assumption}), the spatial modes include in $\boldsymbol{L}\boldsymbol{\Phi}$, and  the normalised spatial modes are obtained as:
\begin{equation}
    \boldsymbol{u}_{m} = (\boldsymbol{L}\boldsymbol{\Phi}\boldsymbol{\Lambda}_\Phi^{-1})_{[:,m]}
    \label{eqs:spatial_modes}
\end{equation}
where the ${\boldsymbol{\Lambda}}_\Phi$ is the diagonal matrix formed by the norm of the column vectors of the matrix $\boldsymbol{L}\boldsymbol{\Phi}$. The subscript $[:,m]$ denotes the $m$-th column of the matrix.
And the amplitude of the modes is:
\begin{equation}
    a_m=|(\boldsymbol{\Lambda}_\Phi\boldsymbol{\Phi}^{-1} \tilde{\boldsymbol{x}}_{1})_{[m]}|
    \label{eqs:mode_amplitude}
\end{equation}
where subscript $[m]$ denotes the $m$-th element of the vector.
Summarizing, we now have $\boldsymbol{u}_{m}$, $a_m$, and $\lambda_m$ to utilise Eq.(\ref{eqs:dmd_assumption}) for predictions of future $x(t_k)$ time steps from any initial values $\tilde{\boldsymbol{x}}_{1}$.

% Participation factor of variable $i$ in mode $m$ is calculated by the element product of $(\boldsymbol{L}\boldsymbol{\Phi})_{i,m} (\boldsymbol{\Phi}^{-1} \tilde{\boldsymbol{x}}_{0})_m$.

\section{Experiments}

Next, we validate the proposed method on two benchmark systems: the IEEE 14-bus system and the WECC (the 179-bus Western Electric Coordinating Council) test case, comprising 5 and 29 generators respectively. These systems, characterized by distinct damping characteristics, enable a comprehensive evaluation of the HODMD method under different conditions. The IEEE 14-bus system requires relatively short temporal windows (10 seconds per trajectory) to capture the rapidly decaying dynamics. In contrast, the WECC system, with less damping and pronounced inter-area oscillations, requires longer observation periods (approximately 20 seconds per trajectory) to accurately capture the oscillation modes, especially the global low-frequency components.

The numerical simulation is conducted using the ANDES Python package \cite{cui_hybrid_2021}. System frequency perturbations are simulated through temporary short-circuit faults applied to individual buses, with a duration of 0.1 seconds ($t$=1.0-1.1s), excluding severe faults leading to system instability. For the IEEE 14 system, we generate a total of 11 trajectories, of which 9 trajectories are training data and the remaining 2 are test data. In the WECC system, we generate 99 trajectories in total, of which 90 are training trajectories and the remaining 9 are test trajectories.

There are two hyperparameters in the HODMD method, the number of truncated modes $r$ and the order $d$. Other studies already discussed how to choose the number of modes $r$ appropriately \cite{jones_application_2022}, and in this paper, an optimal approach is used to choose it\cite{jovanovic_sparsity-promoting_2014}. For the order $d$, we discuss it in the next subsection.

\subsection{Choosing order d}

\begin{figure}[htbp]
\setlength{\abovecaptionskip}{0.cm}
\subfigure
{
    \begin{minipage}[b]{.45\linewidth}
        \centering
        \setlength{\abovecaptionskip}{0.cm}
        \includegraphics[scale=0.85]{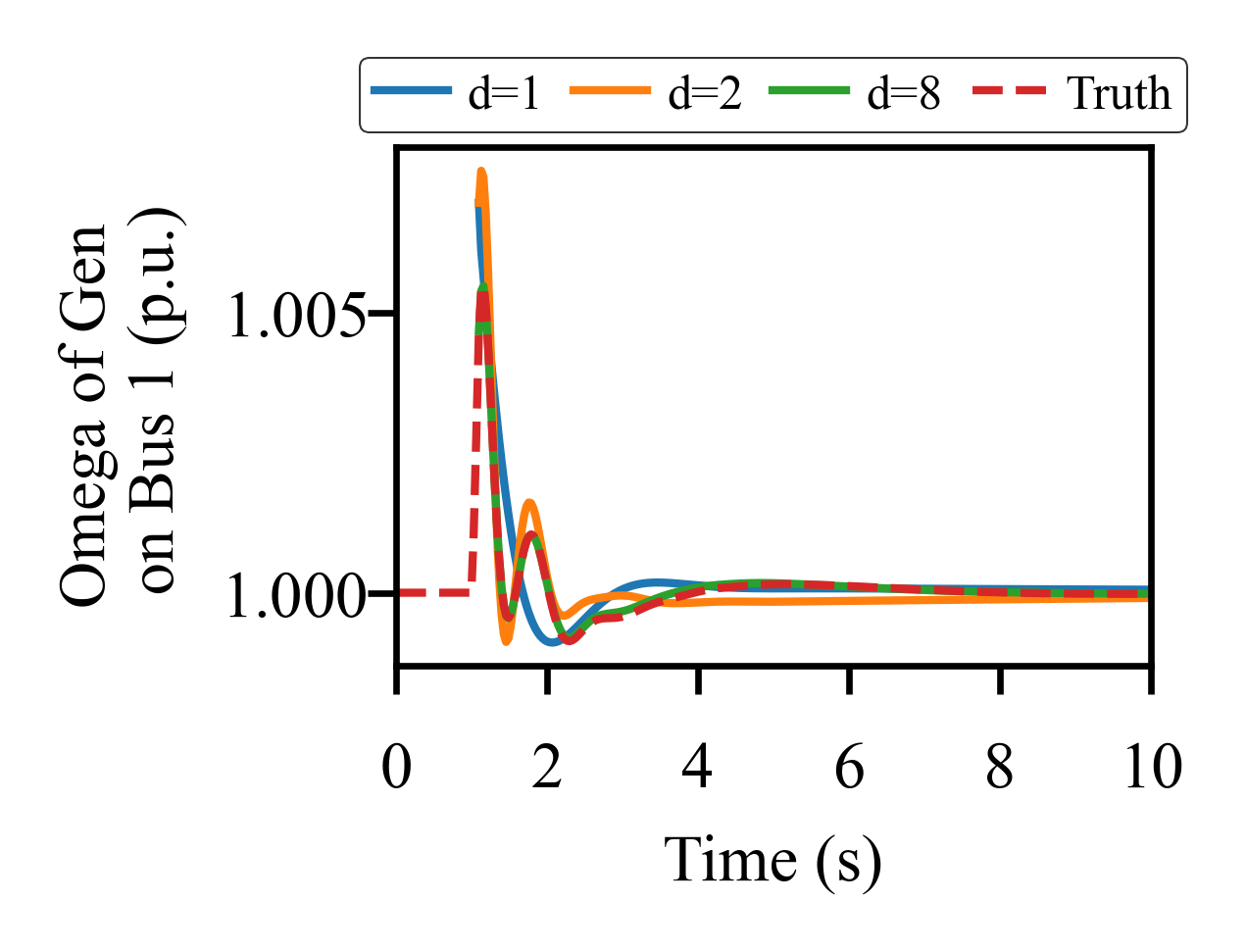}
    \end{minipage}
    % \label{figs:ieee_d(a)}
}
\subfigure
{
    \begin{minipage}[b]{.3\linewidth}
        \centering
        \setlength{\abovecaptionskip}{0.cm}
        \includegraphics[scale=0.85]{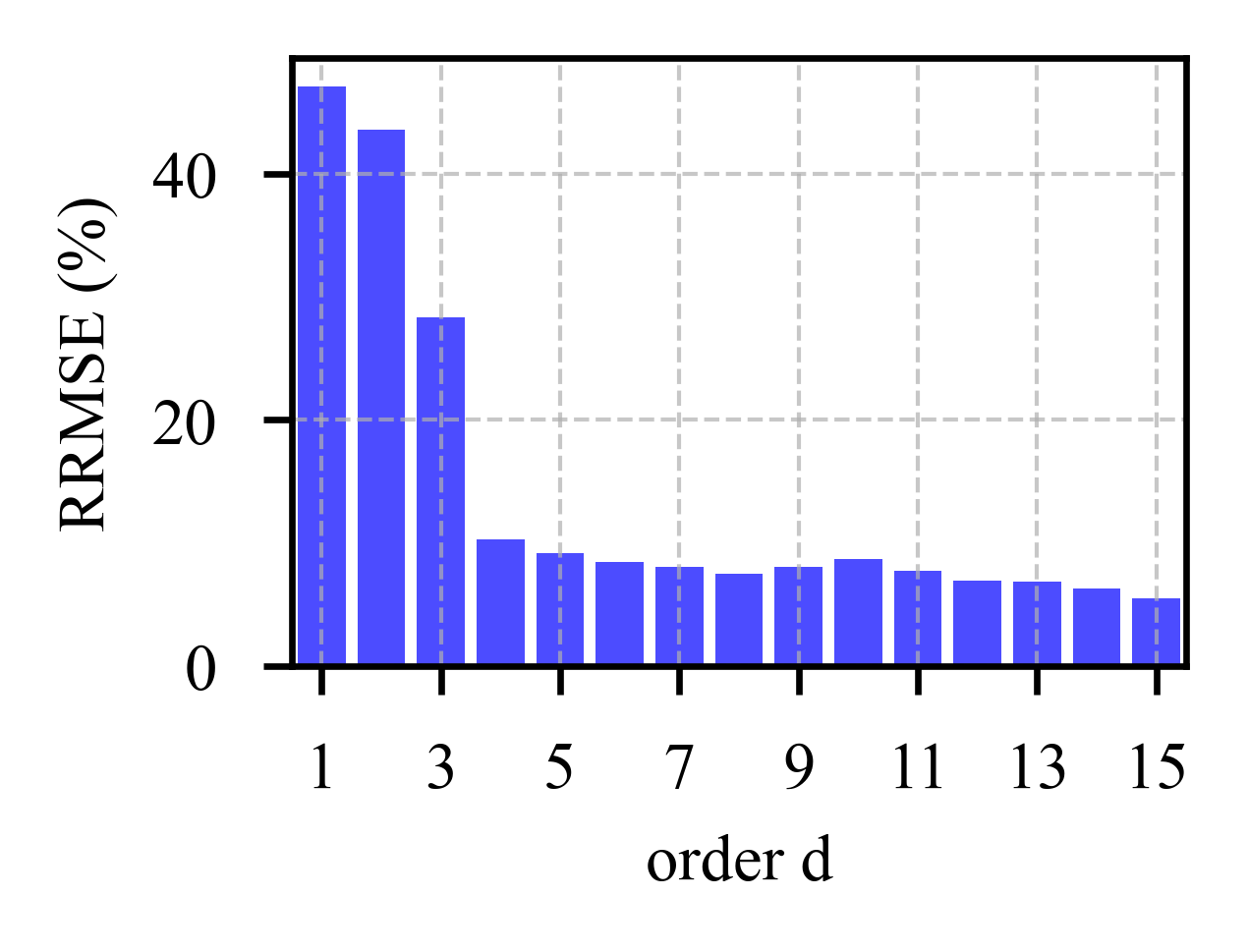}
    \end{minipage}
}
\caption{Comparison of different order d (IEEE 14-bus system).}
\label{figs:ieee_d}
\end{figure}

\begin{figure}[htbp]
\setlength{\abovecaptionskip}{0.cm}
\subfigure
{
    \begin{minipage}[b]{.45\linewidth}
        \centering
        \setlength{\abovecaptionskip}{0.cm}
        \includegraphics[scale=0.85]{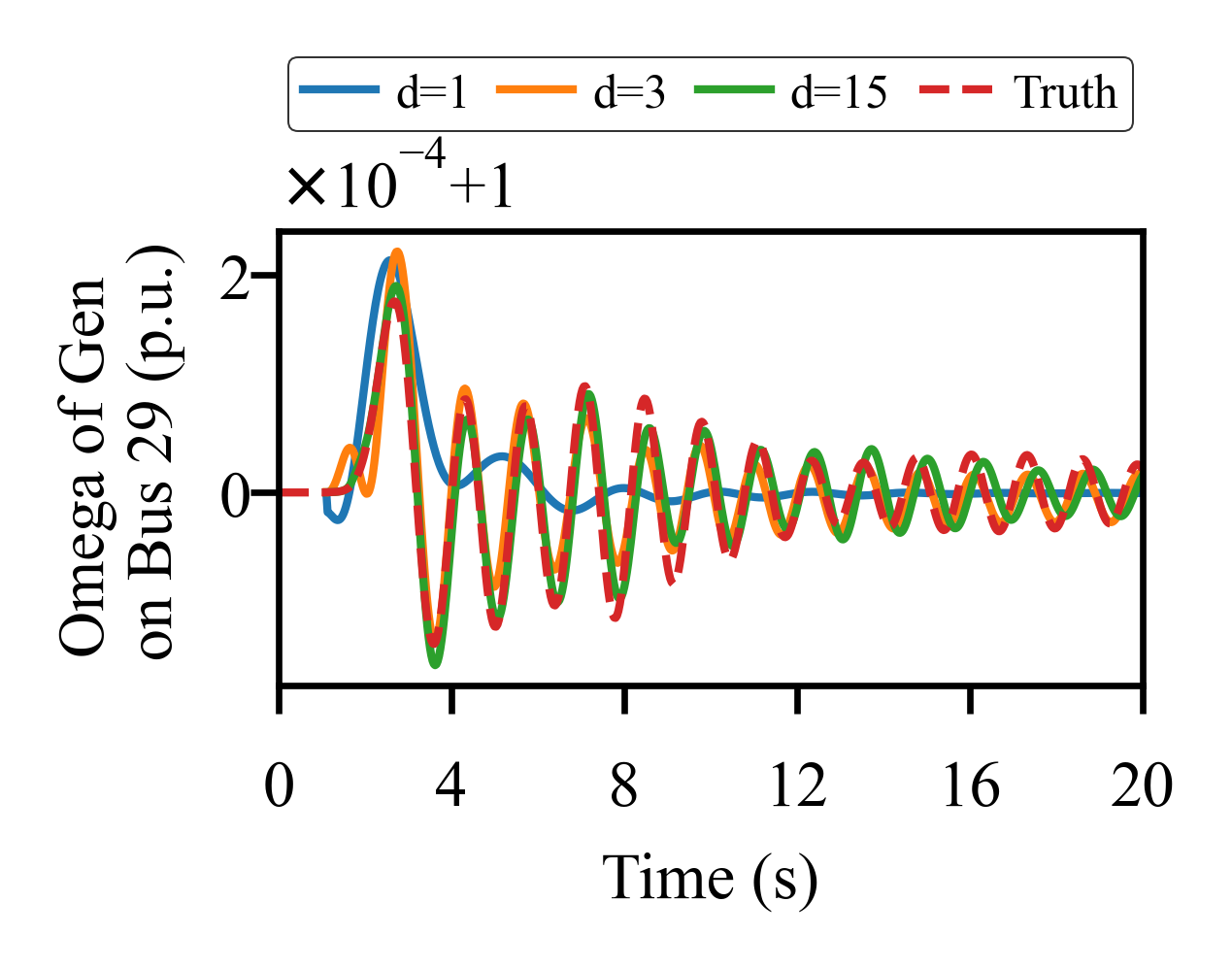}
    \end{minipage}
}
\subfigure
{
    \begin{minipage}[b]{.3\linewidth}
        \centering
        \setlength{\abovecaptionskip}{0.cm}
        \includegraphics[scale=0.85]{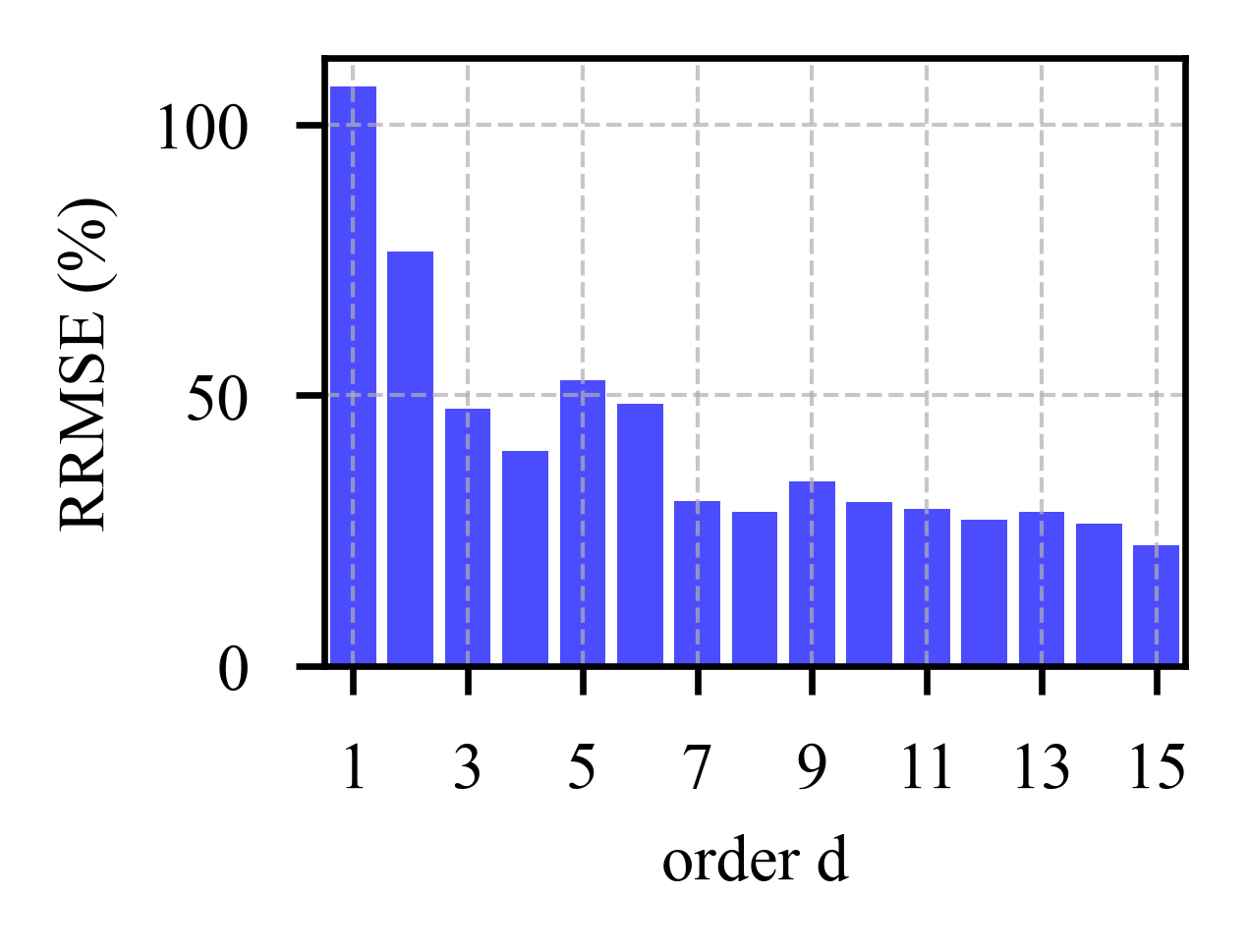}
    \end{minipage}
}
\caption{Comparison of different order d (WECC system).}
\label{figs:wecc_d}
\end{figure}

We choose the hyperparameters order $d$ from the performance of the model on training data. The Relative Root Mean Square Error (RRMSE) is used as the metric, which is suggested in \cite{vega_higher_2020}. The comparison of model performance using different order $d$ is shown in Fig.\ref{figs:ieee_d} and Fig.\ref{figs:wecc_d}.
\begin{equation}
    \mathrm{RRMSE} \equiv 
    \sqrt{
    \frac{
        \sum_{k=1}^{T_\text{total}}\left\|\boldsymbol{x}_{k}-\boldsymbol{x}_{k}^{\text{HODMD }}\right\|_{F}^{2}
    }
    {
        \sum_{k=1}^{T_\text{total}}\left\|\boldsymbol{x}_{k} - \bar{\boldsymbol{x}} \right\|_{F}^{2}}
    }
\label{eqs:rrmse_metrics}
\end{equation}
where $\boldsymbol{x}_{k}^{\text{HODMD}}$ is the reconstruction from HODMD using the procedure introduced in Eq.(\ref{eqs:hodmd_reconstructure}) and $\bar{\boldsymbol{x}}$ is the steady state value in order to remove the dominated constant component, otherwise all the error values become very small. 
% \vspace{-0.7cm}

Fig.\ref{figs:ieee_d} demonstrates that the error decreases as $d$ increases. For $d\geq4$, this reduction in RRMSE becomes marginal. The error at $d=8$ reaches a sufficiently low level, making it an optimal choice. Thus, $d=8$ is chosen. However, the RRMSE is not sufficient to evaluate the oscillatory decay curve fitting quality. At $d=1$, the standard DMD only capture the general trend. It fails to identify the decaying oscillatory frequencies and yields the highest error. When $d=2$, the HODMD prediction closely matches the simulated curve's oscillatory decay characteristics. Yet, the RRMSE values between $d=1$ and $d=2$ show no significant difference. Similar results are observed for $\omega$ of generators at other buses (2, 3, 6, 8) besides Bus 1. Therefore, the RRMSE comparison is preferred over individual RRMSE values.

Fig.\ref{figs:wecc_d} shows that the standard DMD ($d=1$) fails to capture frequency oscillations in the WECC system. The frequency oscillations can be learned when $d\geq3$. Despite relatively high RRMSE values shown in Fig.\ref{figs:wecc_d}, the predicted curve demonstrates good overall fitting quality.  
For oscillatory curves, incorporating more modes generally improves fitting accuracy. The chosen value of 15 remains well below the one-tenth of the amount of data pints recommended by \cite{vega_higher_2020}. Additionally, the sparsity-promoting DMD algorithm \cite{jovanovic_sparsity-promoting_2014} ensures stability and prevents over-fitting. Based on these considerations, we adopt $d=15$ to achieve minimal errors.

\subsection{Performance}

The HODMD method is fitted on the training dataset and then tested on the test dataset. We also compare the HODMD with LANDO \cite{baddoo_kernel_2022} and SINDy \cite{brunton_discovering_2016}. LANDO is a kernel-technique version of the DMD method developed for strongly nonlinear systems, but its performance is sensitive to the form of the kernel function and the choice of parameters. The SINDy method is suitable for both linear or nonlinear methods but is not well suited for high-dimensional systems. 

\begin{figure}[htbp]
\setlength{\abovecaptionskip}{0.cm}
\subfigure
{
    \begin{minipage}[b]{.45\linewidth}
        \centering
        \includegraphics[scale=0.85]{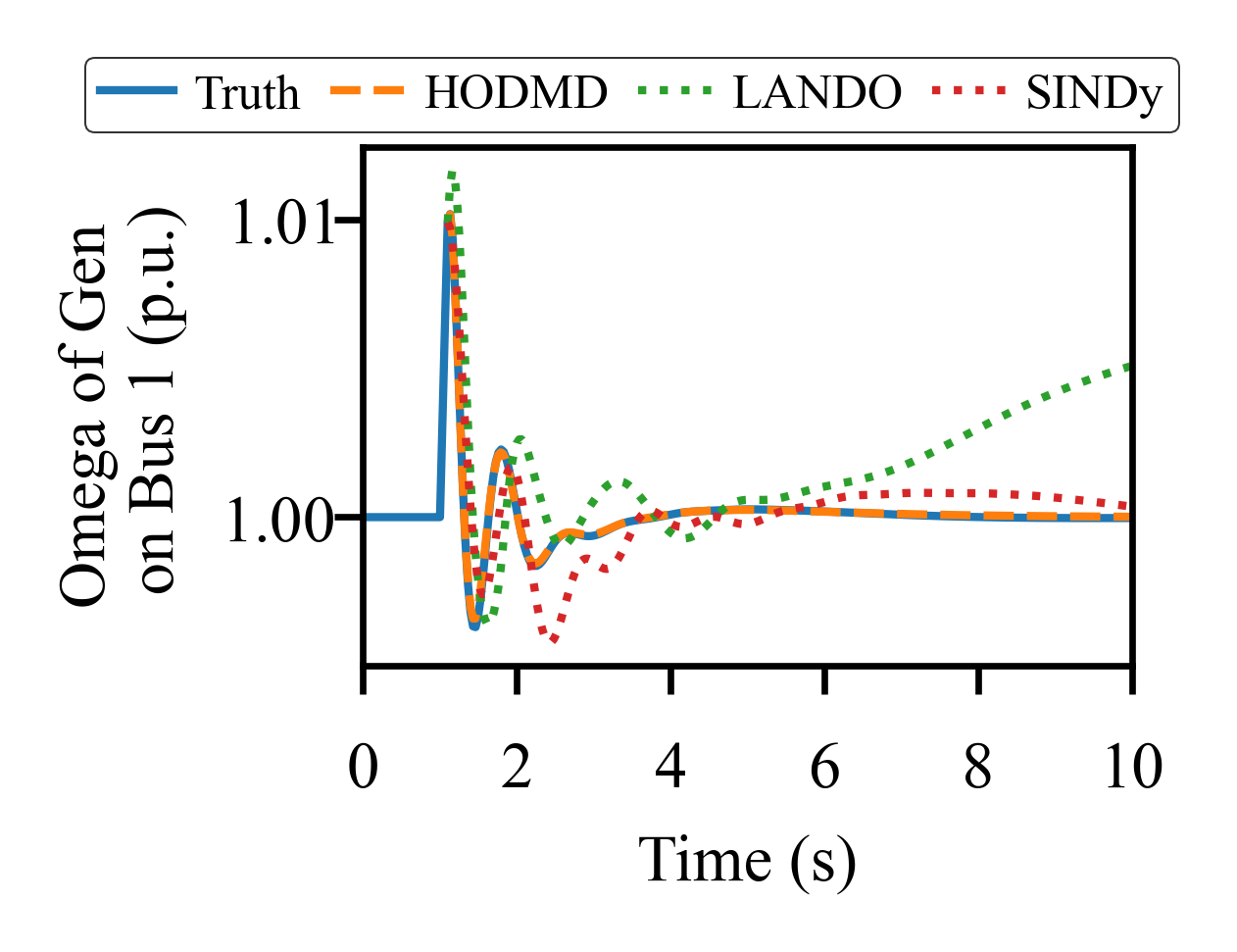}
    \end{minipage}
}
\subfigure
{
    \begin{minipage}[b]{.45\linewidth}
        \centering
        \includegraphics[scale=0.85]{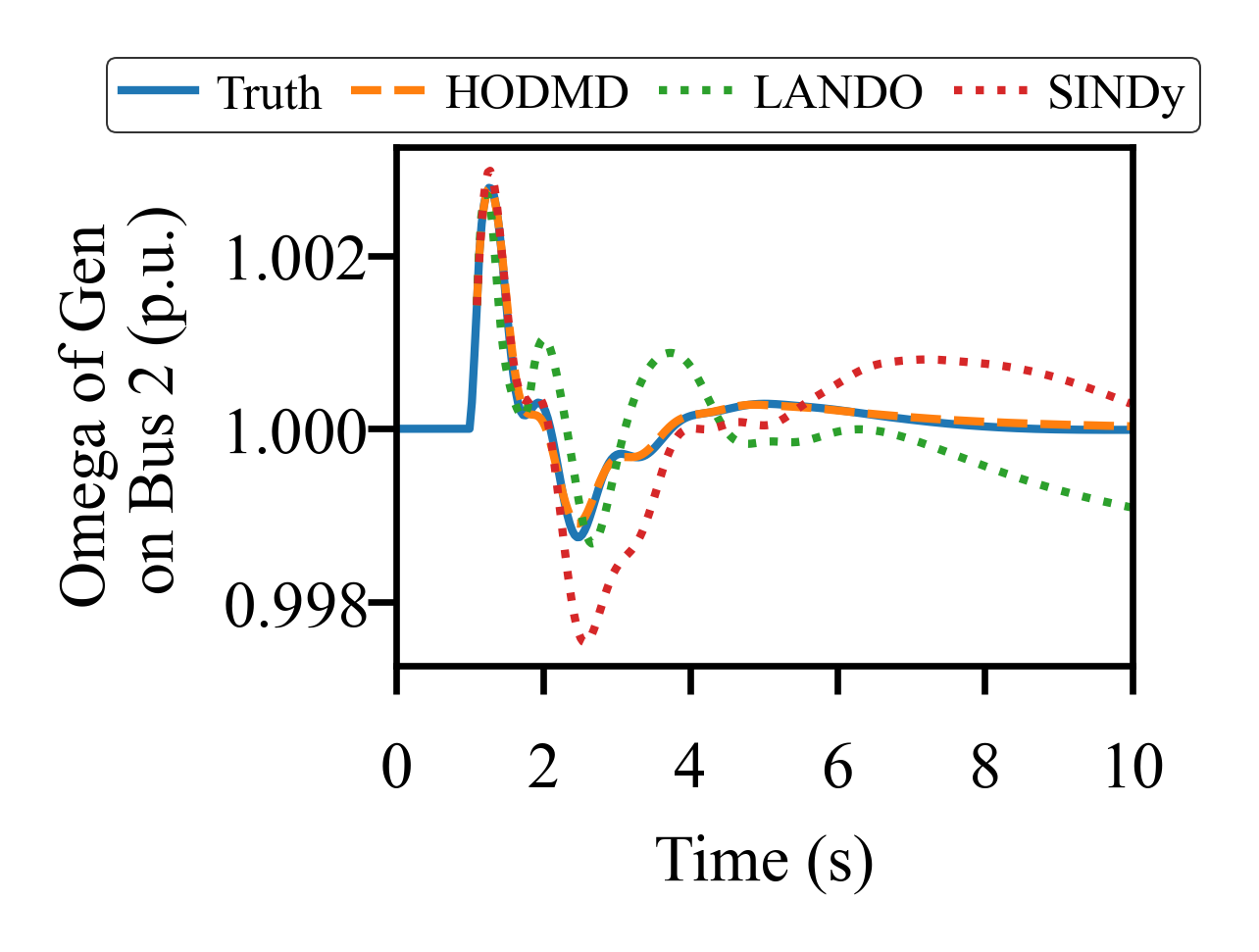}
    \end{minipage}
}
\caption{Performance on tes data ($d=8$ for IEEE 14-bus system).}
\label{figs:ieee14_compare}
\end{figure}

\begin{figure}[htbp]
\setlength{\abovecaptionskip}{0.cm}
\subfigure
{
    \begin{minipage}[b]{.45\linewidth}
        \centering
        \includegraphics[scale=0.85]{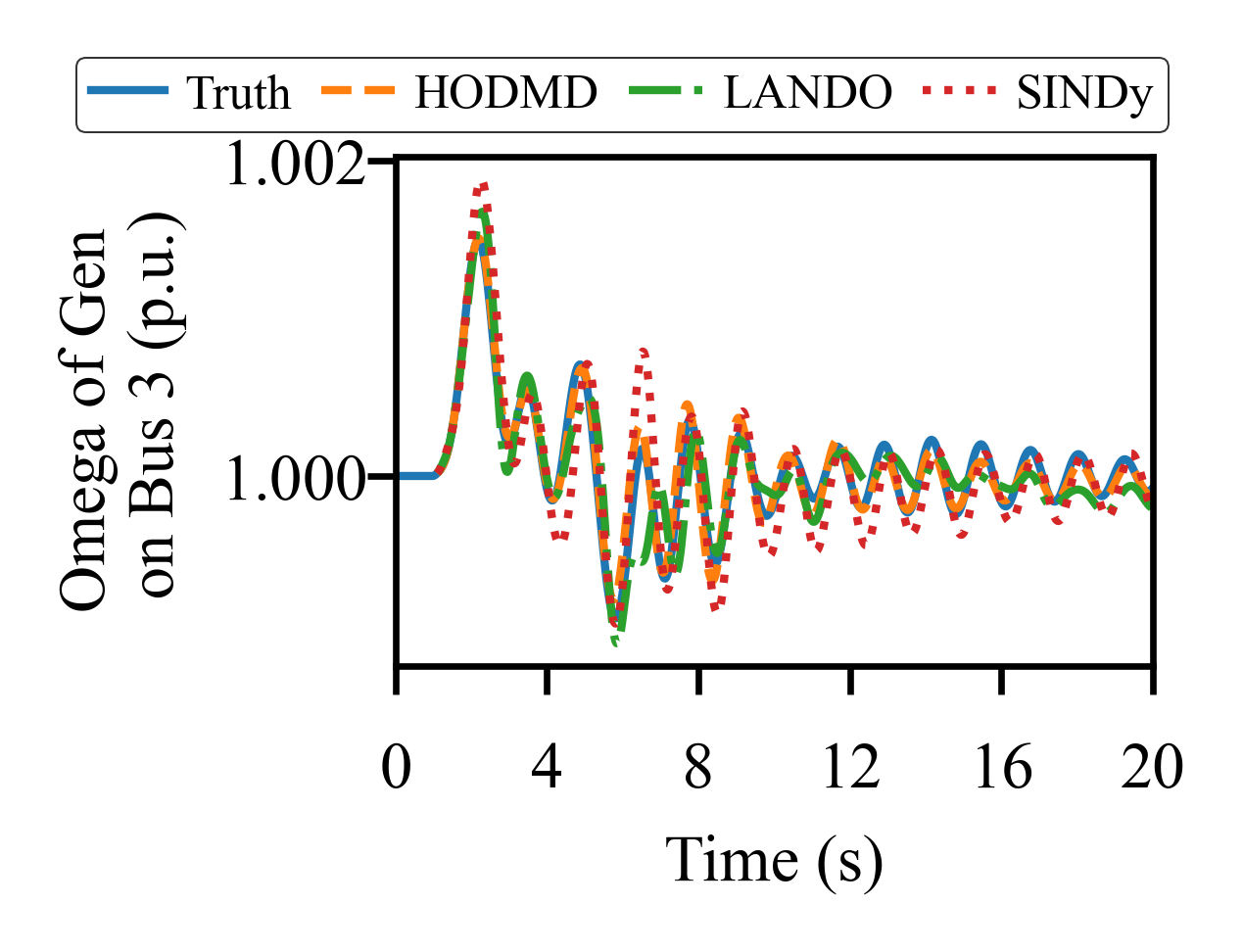}
    \end{minipage}
}
\subfigure
{
    \begin{minipage}[b]{.45\linewidth}
        \centering
        \includegraphics[scale=0.85]{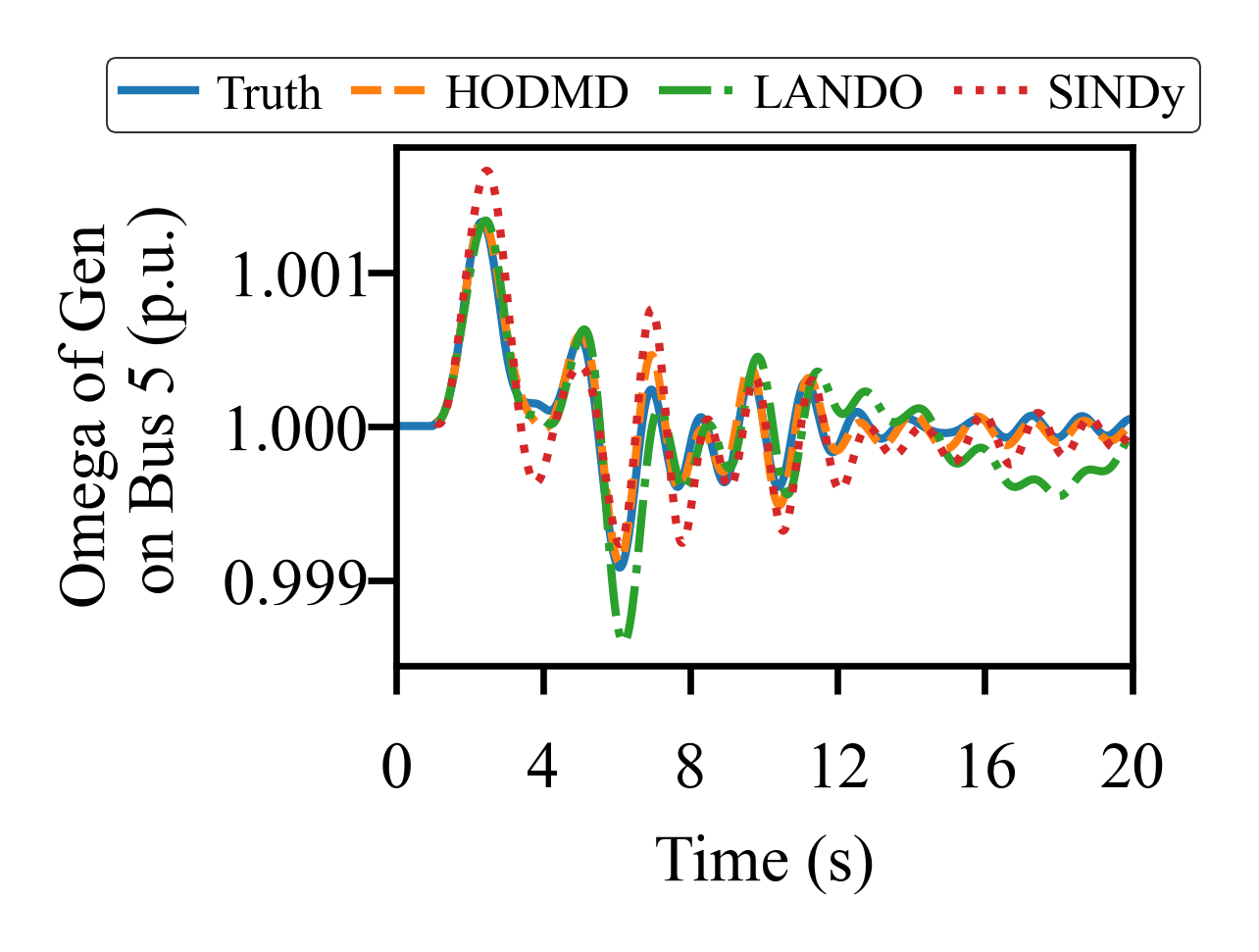}
    \end{minipage}
}
\caption{Performance on test data ($d=15$ for WECC system).}
\label{figs:wecc_compare}
\end{figure}

Fig.\ref{figs:ieee14_compare} and Fig.\ref{figs:wecc_compare} show the prediction curves of these three methods for the test data. After iterative adjustments, a Gaussian kernel with the best performing length scale parameter is used in LANDO and the rank of its SVD is $r=10$, because the higher rank leads to unstable prediction results. the basis function of SINDy is only linear because any higher-order polynomial eigenfunctions or trigonometric functions lead to divergent prediction curves. 
From these figures, HODMD fits best. The RRMSEs of HODMD, LANDO, and SINDy on test data of the IEEE 14-bus system are 15.2\%, 71.9\% and 71.9\%, and on test data of the WECC system are 36.7\%, 42.1 \%, and 37.1\%. Therefore, HODMD is more applicable to learning the frequency dynamics of power systems, i.e., LANDO and SINDy methods do not apply to weakly nonlinear high dimensional systems.

\vspace{-0.7cm}

\subsection{Robust to noise}

\begin{figure}[htbp]
\setlength{\abovecaptionskip}{0.cm}
\centerline{\includegraphics[scale=0.85]{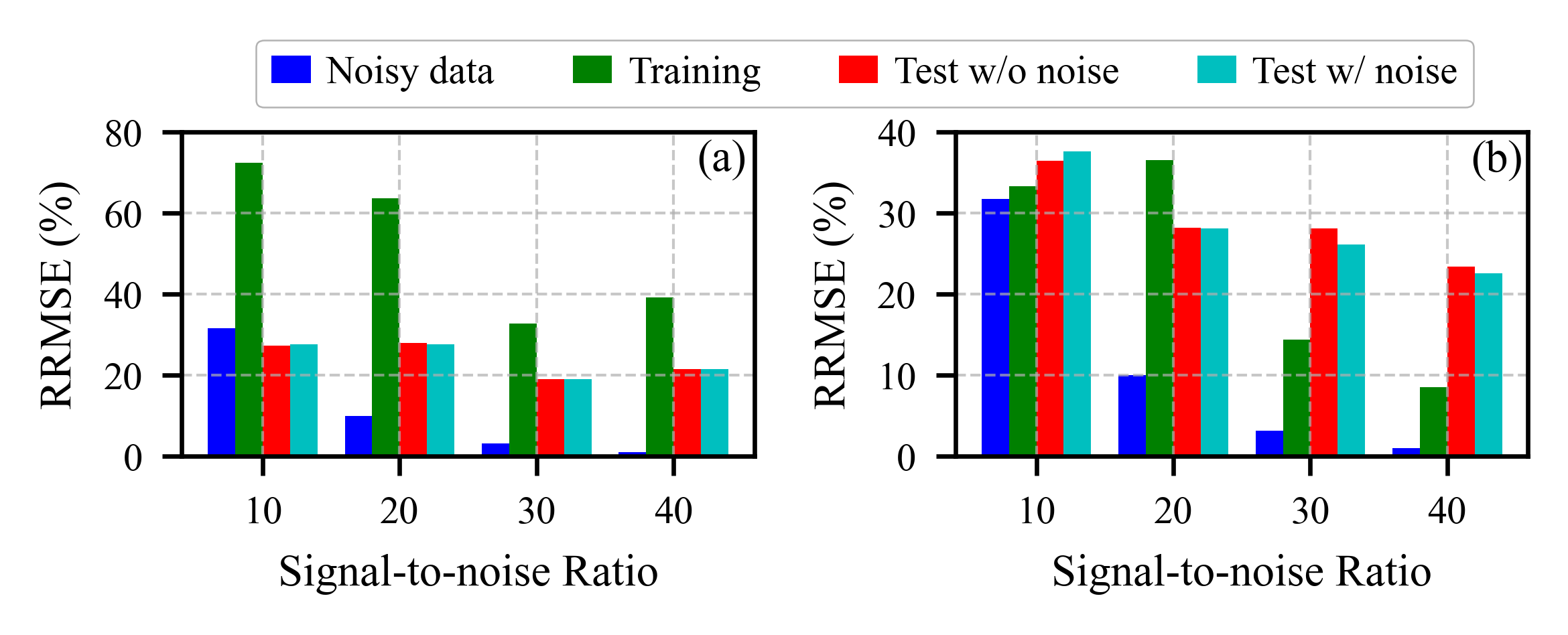}}
\caption{RRMES with noisy data: (a) IEEE 14-bus system; (b) WECC system.}
\label{figs:snr_rrmse} 
\end{figure}

\begin{figure}[htbp]
\setlength{\abovecaptionskip}{0.cm}
\centerline{\includegraphics[scale=0.88]{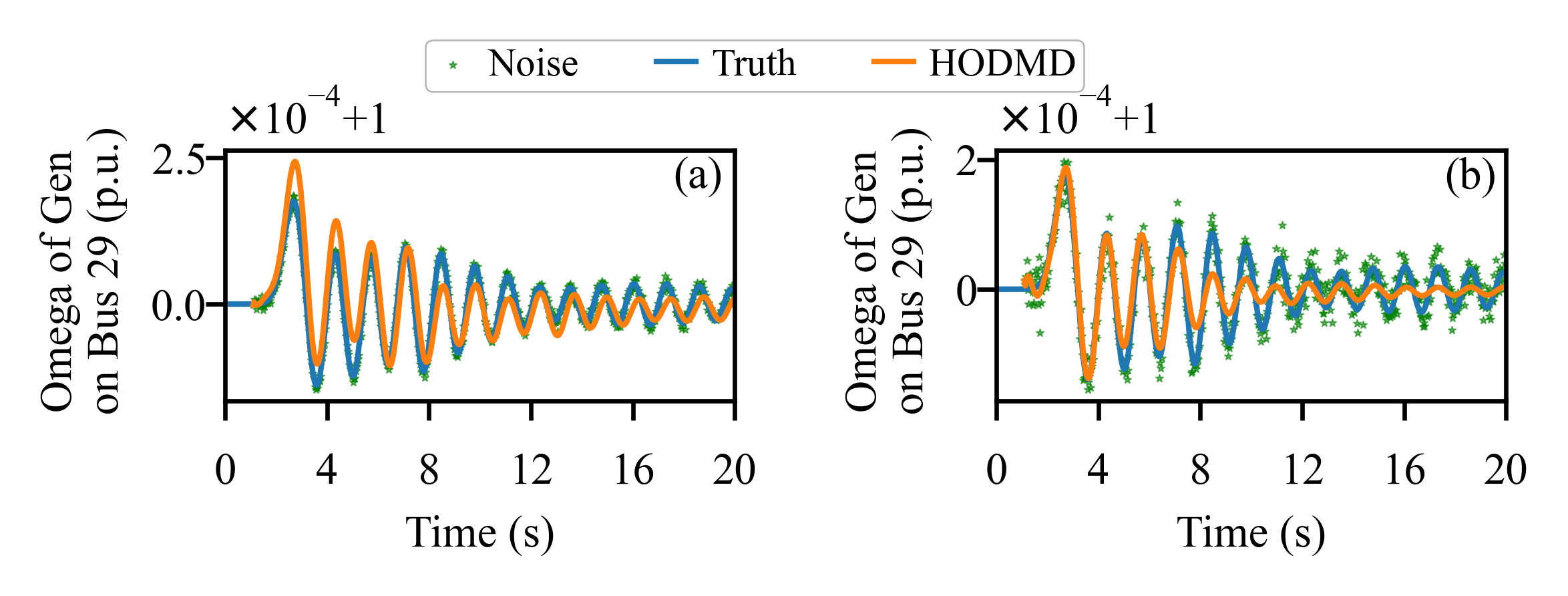}}
\caption{Training with noise on WECC system: (a) SNR 20; (b) SNR 10.}
\label{figs:wecc_noise_training} 
\end{figure}

\begin{figure}[htbp]
\setlength{\abovecaptionskip}{0.cm}
\centerline{\includegraphics[scale=0.88]{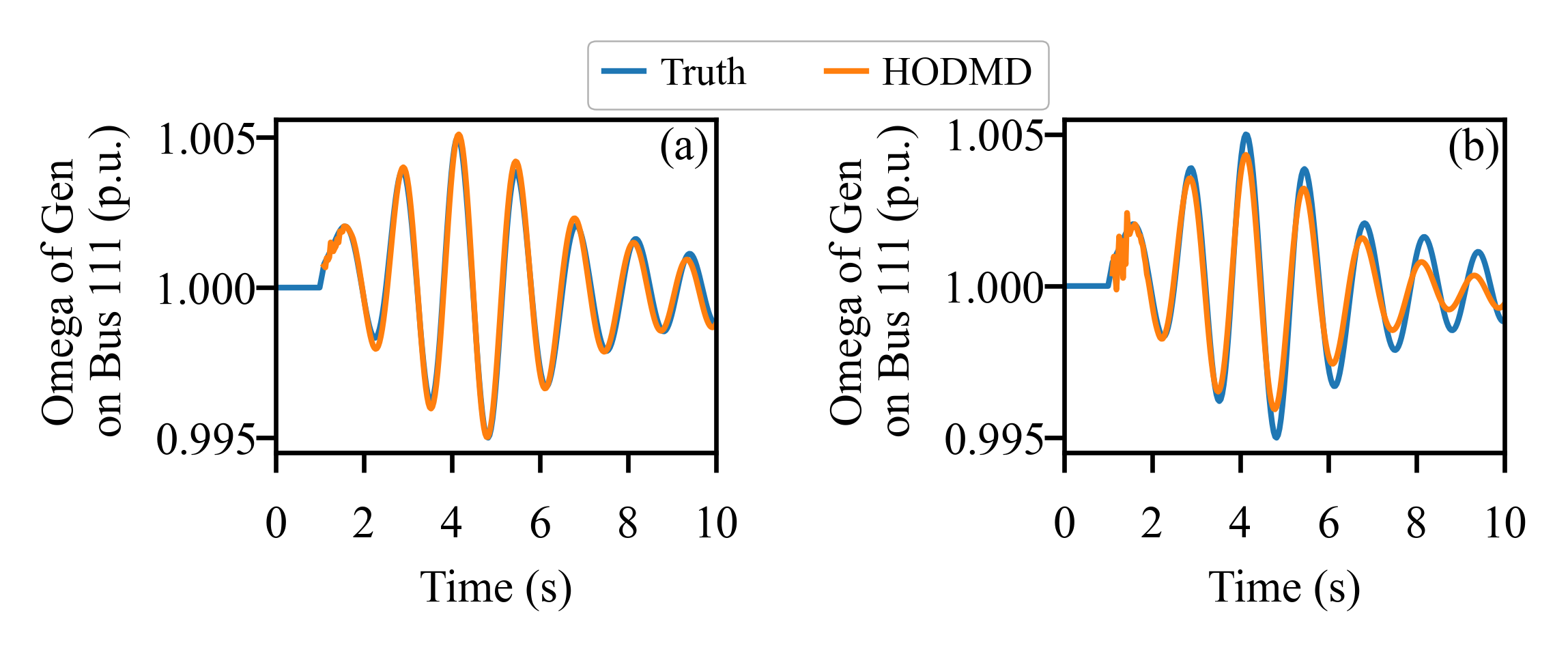}}
\caption{Prediction with noise on WECC system: (a) SNR 20; (b) SNR 10.}
\label{figs:wecc_noise_prediction} 
\end{figure}

Although we don't explicitly consider noisy measurements, HODMD is robust to noise. To demonstrate how HODMD could handle noisy measurements, e.g. due to renewable generators, we add a small Gaussian noise to both training and test data. The mean of the noise is the steady-state value of the variable and the variance of the noise is measured by the signal-to-noise ratio (SNR). As the SNR decreases, the noise increases.
The RRMSE values of the noisy data compared to the original data without noise are about 10\% and 31\% for SNR of 20 and 10, respectively, as shown in Fig.\ref{figs:snr_rrmse}. The performance of this method for noisy data is similar in both systems: as the SNR decreases, the RRMSE on the training data increases significantly but remains smaller than the training error of the standard DMD in the absence of noise. The range of variation of the error on the test data is yet smaller.
The curves for the training data under noise and the predicted curves, respectively, are shown in Figs.\ref{figs:wecc_noise_training}, \ref{figs:wecc_noise_prediction}. Except at the oscillatory decay where the noise drowns out the signal, the other parts can be fitted very well and are not even deteriorated by the increase in noise. Thus, HODMD is robust to noise in terms of learning frequency dynamics.

% \vspace{-0.7cm}

% \vspace{-0.7cm}

\subsection{Modal analysis}

\begin{figure}[htbp]
\setlength{\abovecaptionskip}{0.cm}
\centerline{\includegraphics{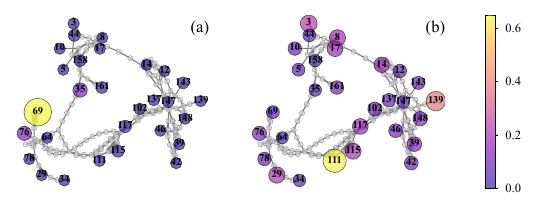}}
\caption{Local spatial mode in WECC system:(a) Mode 260; (b) Mode 288.}
\label{figs:wecc_local_spatial_mode}
\end{figure}

\begin{figure}[htbp]
\setlength{\abovecaptionskip}{0.cm}
\centerline{\includegraphics{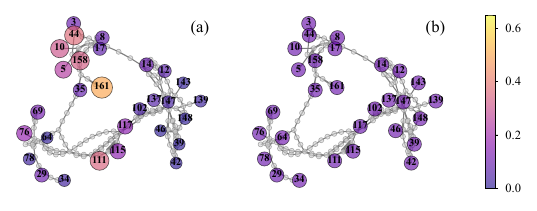}}
\caption{Global spatial mode in WECC system:(a) Mode 298; (b) Mode 336.}
\label{figs:wecc_global_spatial_mode}
\end{figure}

Using the spatial patterns and eigenvalues of HODMD, the global and local frequency dynamics of the system can be revealed.
Truncated modes number $r=380$ are obtained by the optimal SVD for HODMD of order $d = 15$. The spatial modes of this system are calculated using Eq.(\ref{eqs:spatial_modes}). Examples of the two local modes are shown in figure \ref{figs:wecc_local_spatial_mode}, where the colour values are the corresponding elements in $\boldsymbol{u}_{m}$, the bus numbers connecting with generators as labelled on the figure, and the other buses are plotted as small grey circles. Modes 260 and 288 are the local dynamics of the generator at Bus 111 and Bus 69 respectively. Modes 298 and 336 as shown in Fig. \ref{figs:wecc_global_spatial_mode} are global modes because the spatial modes do not concentrate on a few nodes, especially Mode 336.

% \vspace{-0.7cm}

We verify these modes by analysing the frequency response curves. An example is taken from the test data. When the disturbance occurs at Bus 70, which is close to Bus 69, the initial amplitude am of the modes is calculated by eq.(\ref{eqs:mode_amplitude}). The first 30 modes with the largest amplitude are filtered out. These include modes 260, 288, 298, 329, and 336, while the other modes have larger decay coefficients so they quickly decay to 0 on the time domain curve. Mode 329 is not oscillatory, but a monotonically decaying mode with a smaller decay coefficient, which is also a global spatial mode. Details are in Table \ref{tab:wecc_modes}.

\begin{table}
    \centering
    \setlength{\abovecaptionskip}{0.cm}
    \caption{Modes of HODMD}
    \begin{tabular}{cccccc}
        \hline
        Mode & frequency(Hz) & decay(p.u.)& damping(\%)& L/G\textsuperscript{a} & Bus\textsuperscript{b}\\
        \hline
        260  & 1.26          &0.24   & 3.05       & L   & 69\\
        \hline
        288  & 0.77          &0.17   & 3.59       & L   & 111\\
        \hline
        298  & 0.68          &0.19   & 4.53       & G   & -\\
        \hline
        336  & 0.10          &0.39   & 51.94      & G   & -\\
        \hline
        329  & -             &0.03   & -          & G   & -\\
        \hline
        \multicolumn{6}{p{0.9\linewidth}}{\raggedright \textsuperscript{a}: Local or global spatial mode. \textsuperscript{b}: To which generator at the bus the local modes are related.} \\
    \end{tabular}
    \label{tab:wecc_modes}
\end{table}

\begin{figure}[htbp]
\centering
\setlength{\abovecaptionskip}{0.cm}
\includegraphics{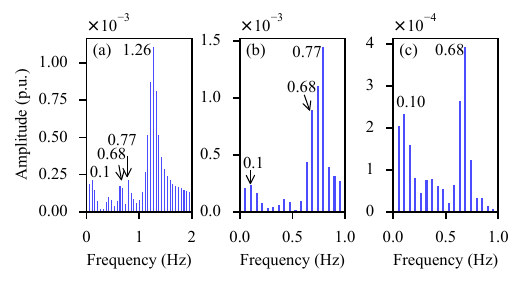}
\caption{Fast Fourier Decomposition on the $\omega$ trajectory of Gen at Bus:(a) 69; (b) 111; (c) 161. (b) and (c) have almost 0 components after 1 Hz.}
\label{figs:fft_wecc}
\end{figure}
% \vspace{-0.5cm}

Fast Fourier analysis is conducted on simulated normalized angular frequency profiles at Buses 69, 111, and 161 (Fig.\ref{figs:fft_wecc}). The analysis also revealed two distinct categories of modes: local and global spatial modes. For local modes, Bus 69 exhibited a dominant frequency component at 1.26 Hz corresponding to Mode 260, while Bus 111 showed a maximum component at 0.77 Hz aligning with Mode 288. These correlations validate HODMD's capability in correctly identifying local spatial modes. Regarding global spatial modes, all profiles demonstrated a common low-frequency peak at 0.1 Hz (Mode 336) and a secondary peak at 0.68 Hz (Mode 298), though the latter was less distinct in Bus 111's profile due to its proximity to the local mode frequency of 0.77 Hz.
The observed frequency distribution, where global modes exhibit lower frequencies compared to local modes, aligns with established power system oscillation characteristics. This frequency pattern validates both the HODMD-identified modes and demonstrates the method's superior capability in capturing temporal-spatial mode interactions, an advantage over traditional DMD approaches.

\section{Discussion}

This study advances this field in data-driven learning of frequency dynamics. By incorporating delayed embedding techniques from HODMD, our work extends the traditional modal decomposition framework \cite{barocio_dynamic_2015} beyond linear analysis into the nonlinear domain—an approach that has received limited attention in existing research. The method's distinctive advantage lies in its independence from basis function selection, thereby circumventing the a priori knowledge requirements that often constrain other nonlinear approaches, like SINDy and LANDO. Furthermore, through SVD-based dimensionality reduction, the methodology maintains computational efficiency even when applied to high-dimensional systems by projecting state variables onto a lower-dimensional subspace.

However, the research in this paper also has some limitations at present, for example, this research is currently based only on simulation data, as the measurement data still needs data pre-processing and some additional techniques for dealing with nonlinear dynamics to achieve good results, which is also our next goal.

\section{Conclusion}

This paper presents an efficient technique based on the HODMD method for learning the frequency dynamics of power systems. This approach is both a natural extension of standard DMD methods, robust to noise, able to handle large amounts of data in high dimensions and able to perform global and local modal analysis. 
Compared to linear methods, such as the standard DMD method, this result shows that the accurate learning and prediction of frequency dynamics requires the introduction of nonlinear properties. HODMD, which is simple in form and powerful enough, is more suitable for learning frequency dynamics in power systems than other nonlinear methods and has the potential for real-world application.

\section*{Acknowledgment}
Xiao Li thanks Guilherme Silva for his thought-provoking discussion.


\end{document}